\documentclass[aps,pre,showpacs,showkeys,onecolumn,10pt]{revtex4-1}
\usepackage[dvips]{graphicx}
\usepackage[dvips]{color}
\usepackage{hyperref,breakurl,amsmath,amssymb,pst-node,eepic}
\usepackage{bbold}

\begin{document}

\title{Dynamics of Wealth Inequality}

\author{Zdzislaw Burda}
\email{zdzislaw.burda@agh.edu.pl}
\affiliation{Faculty of Physics and Applied Computer Science, AGH University of Science and Technology,  al. Mickiewicza 30, PL--30059 Krak{\'o}w, Poland}
\author{Pawel Wojcieszak}
\email{pawlo1604@tlen.pl}
\affiliation{Faculty of Physics and Applied Computer Science, AGH University of Science and Technology,  al. Mickiewicza 30, PL--30059 Krak{\'o}w, Poland}
\author{Konrad Zuchniak}
\email{konrad.zuchniak@gmail.com}
\affiliation{Faculty of Physics and Applied Computer Science, AGH University of Science and Technology,  al. Mickiewicza 30, PL--30059 Krak{\'o}w, Poland}

\begin{abstract}
We study an agent-based model of evolution of wealth distribution in a macro-economic system. The evolution is driven by multiplicative stochastic fluctuations governed by the law of proportionate growth and interactions between agents. We are mainly interested in interactions increasing wealth inequality that is in a local implementation of the accumulated advantage principle. Such interactions destabilise the system. They are confronted in the model with a global regulatory mechanism which reduces wealth inequality. There are different scenarios emerging as a net effect of these two competing mechanisms. When the effect of the global regulation (economic interventionism) is too weak the system is unstable and it never reaches equilibrium. When the effect is sufficiently strong the system evolves towards a limiting stationary distribution with a Pareto tail. In between there is a critical phase. In this phase the system may evolve towards a steady state with a multimodal wealth distribution. The corresponding cumulative density function has a characteristic stairway pattern which reflects the effect of economic stratification. The stairs represent wealth levels of economic classes separated by wealth gaps. As we show, the pattern is typical for macro-economic systems with a limited economic freedom. One can find such a multimodal pattern in empirical data, for instance, in the highest percentile of wealth distribution for the population in urban areas of China.  

\end{abstract}


\maketitle

\section{Introduction}
Wealth inequality in the world has been continuously rising since the seventies of the last century \cite{p,sz}. According to Thomas Piketty, the author of the book {\em Capital in the Twenty-First Century} \cite{p,p1}, this process is an inherent feature of global capitalism and will dominate the world economy and social structure in the twenty first century unless some special measures are taken to inhibit or reverse it. Such measures require a very strong international integration and implementation of economic interventionism in the scale of the whole globe. Left to itself the growth of wealth inequality may at some point threaten the democratic order and lead to strong instabilities \cite{p,ox}. The scale of wealth inequality is well illustrated by data on wealth concentration \cite{ox,cs}. The wealth of the richest one percent of the world population is roughly equal to the wealth of the remaining $99\%$.  
The number of richest individuals who own the same wealth as the poorer half of the world  decreases year by year. According to Oxfam's estimates based on  Credit Suisse Global Wealth Databook this number was equal eight at the end of 2016. Even though it is only an estimate, which may be slightly biased, it certainly reflects an enormous scale of wealth concentration and thus also of wealth inequality.  

Both empirical and theoretical research on wealth distribution is an important and active branch of contemporary economics (see \cite{ds,bb} for review) and it is closely interrelated with the problem of optimal conditions for economic growth (\cite{obt,b}). The modern, systematic analysis of income and wealth distributions dates back to Vilfredo Pareto who made an important observation about wealth and income statistics \cite{p2}. The status of this observation has evolved in the course of time into what is known today as the Pareto law \cite{s}. The law concisely summarises statistical properties of the distribution of income and wealth of the richest part of the population. Another important contribution was made by Robert Gibrat who formulated the law of proportionate effect 
\cite{g} which lies at the heart of economic processes responsible for evolution of wealth distribution viewed from the large macro-economic scale. While studying the issue of wealth inequality one has to go beyond the equilibrium formulation and look at dynamics of the underlying processes from a non-stationary perspective in a non-equilibrium framework. In this framework one can for example analyse stability of wealth distribution with respect to perturbations of tax rates, saving tastes, productivity levels, return rates on capital, growth rates, {\em etc.} In the simplest case \cite{ps} the evolution is modelled by a single stochastic evolution equation. The population statistics can be imitated by replacing parameters of this equation by random variables or associated stochastic processes. For example the return rate can be modelled as a Gaussian variable with a non-zero width, or as a stochastic process whose values are positively correlated with wealth if one wants to model the Matthew effect of accumulated advantage. There is some freedom in incorporating various aspects of population statistics  in this approach, but unfortunately one cannot implement correlations between individuals or their collective behaviour, as for example herding \cite{cb}. To do this one
has to study the whole population of individuals. It can be done in the framework of agent-based modelling which was pioneered in \cite{a}, (see \cite{szsl} for review).
We adopt it here to describe dynamics of wealth distribution in a macro-economic system by a set of interacting stochastic equations. Piketty argues that the primary cause for wealth inequality to rise is that the rate of return on capital is greater than the economy growth rate. Individuals who have a surplus of capital can invest it and multiply it at a rate that is usually larger than the economic growth rate \cite{p,p1,ps}. This is, roughly speaking, equivalent to what is commonly known as the Matthew effect of accumulated advantage, or the rich-get-richer (poor-get-poorer) mechanism, which provides a generic explanation of inequalities and heterogenous behaviour observed in many phenomena throughout many scientific disciplines \cite{y,sim,ba}. In this work we discuss how to incorporate the rich-gets-richer mechanism into the framework of agent-based modelling and discuss how it influences the population dynamics and the evolution of wealth distribution in the macro-economical scale. 

Dynamics of wealth distribution is a complex phenomenon which is determined by many factors including legal regulations, taxes, welfare, innovations, macroeconomic conditions, international relations, education, consumption, the existence of tax havens, intergenerational wealth transfers, inheritance, lifecycle accumulation, dynamics of self-made fortunes and many others \cite{ds,bb}. It is extremely hard to find a quantitative representation of all these factors and to implement them into a realistic model which would be capable of mimicking behaviour of individuals and their decisions. Even if such a model had existed it would have been highly non-linear and would have had thousands of parameters. This is a typical situation which is referred to as a curse of dimensionality, meaning that a model having ambitions to be maximally realistic suffers in fact from over-parameterisation. The model looses its predictive power and fails to unambiguously explain observed features. Empirical data can be explained by different combinations of parameters. Moreover, a little change of a single parameter in one sector of the model may completely change results in other sectors.

An alternative approach is to invoke the idea of reductionism which has a long tradition in natural and economic sciences, in general, and in research on wealth distribution, in particular. We take this approach here. We consider a statistical agent model which describes a network of interacting individuals. The model is a variant of the Bouchaud-M\'{e}zard model \cite{bm}. The main difference as compared to the original model is that we consider interactions amplifying wealth inequality: when two agents interact the richer one gets richer and the poorer one gets poorer as a result of the interaction. This is a local (microeconomic) implementation of the accumulated advantage principle. Obviously, such local forces destabilise the system and drive it out of equilibrium. If there were no other forces the system would be unstable. In many real situations, however, and also in our model some stabilising forces come into being. They are introduced by policy makers as regulatory mechanisms to reduce wealth inequality in the system. In effect the system may evolve to an interesting steady state reflecting these two competing factors. The situation is similar to that known for many systems in statistical and quantum physics. For example a Coulomb gas which consists of repelling particles is unstable. However if one puts it to a confining potential it may form a stabilised state for which Coulomb repulsion is balanced by external forces coming from the potential. The most famous example is the Coulomb gas picture that explains universality classes and highly non-trivial statistics of eigenvalues of random matrices as a net effect of local repulsion and global confinement \cite{d,f}. A less known example of this type is the phase transition triggered by a global constraint in the backgammon model (known also as urn model or balls-in-boxes model) \cite{bbj,bbbj}. Here we look for similar effects in agent-based models.

Research on wealth distribution is a very active branch of science both in economics \cite{sz,p1,ds,xg,dn,ap,bpa,bbs,bb,obt,bfp} and econophysics \cite{b,cb,szsl,bm,ls,ikr,yr,bjn,cccc,mhs,ls1,bss,ls2}. An important contribution of econophysics to this field is the development of multi-agent models which are capable to describe statistical, collective effects and large-scale emergent phenomena. In such models behaviour of a single agent (individual) can be understood only in relation to behaviour of others. Agents shape the system behaviour and the system shapes their behaviour. This feed-back is often responsible for emergence of large-scale phenomena in macro-economic systems. The Bouchaud-M\'{e}zard model that we build on belongs to this category. So before we define our model let us recall some fundamental ideas that underlie it. They are present in many other models on wealth distribution.

\section{Milestones}
\subsection{Gibrat's rule}

The basic idea is that the wealth of an individual changes multiplicatively rather than additively \cite{g}. In the discrete time description one can write an equation for wealth $W_\tau$ at time $\tau$
\begin{equation}
W_\tau = \Lambda_\tau W_{\tau-1} \ ,
\label{prop_eff}
\end{equation}
where $\Lambda_\tau$ is a random factor. By iterating this equation one obtains a stochastic evolution of wealth. The physical time that elapses between two instants of time $\tau-1$ and $\tau$ (indexed by integers) is equal to $\Delta t$ which is a period characteristic of the problem in question. It can be a day, month, year, a fiscal period, or any other suitable time unit. For example in intergenerational wealth transfer issues it is the generation time which is the average time between two consecutive generations. Expressed in physical units, time is $t = \tau \Delta t$. Eq. (\ref{prop_eff}) is commonly known as the law of proportionate effect or Gibrat's rule \cite{g}. If $\Lambda$'s are independent of $W$'s and of each other and the probability distribution of $\Lambda$ does not change in time then Eq. (\ref{prop_eff}) describes one-dimensional geometric Brownian motion in the discrete time formalism. If one takes logarithm of both sides of Eq. (\ref{prop_eff}) one gets an equation $\log W_\tau = r_\tau + \log W_{\tau-1}$, being a discrete version of the standard Brownian motion for $\log W_\tau$ with random changes given by random logarithmic returns $r_\tau = \log \Lambda_\tau$. As follows from the central limit theorem, for a broad class of probability distributions of returns which do not have too heavy tails \cite{gk}, the total return $\log W_\tau/W_0$ behaves in the limit of large time $\tau\rightarrow \infty$, like a normal random variable with the mean proportional to $\tau$ and the standard deviation proportional to $\sqrt{\tau}$. Moreover when the returns $r_\tau$ themselves are normal variables with the mean proportional to the time interval $\Delta t$ {\em i.e.} $\mu=\mu_0 \Delta t$ and the standard deviation proportional to the square root of the time interval {\em i.e.} $\sigma =\sigma_0 \sqrt{\Delta t}$ then one can take the continuous time limit of Eq. (\ref{prop_eff}), $\Delta t \rightarrow 0$. In this limit Eq. (\ref{prop_eff}) transforms to a stochastic differential equation which in It\^{o} calculus \cite{i,vk} has the form $dW = W \left( (\mu_0 + \sigma^2_0/2)dt + \sigma_0 dB(t)\right)$ where $B(t)$ is the Wiener process. The letter $B$ refers to Brownian motion being the most prominent physical realisation of the Wiener process. The distribution of wealth $W_\tau/W_0$ evolves as a log-normal distribution $\log\mathcal{N}(\mu \tau, \sigma^2 \tau)$ (or equivalently in physical time units as $\log\mathcal{N}(\mu_0 t, \sigma^2_0 t)$). The mean value grows (or falls off) exponentially $E(W_\tau) = W_0 e^{\tau \Delta t(\mu + \sigma^2/2)}$. The average wealth per capita 
\begin{equation}
\overline{W}_\tau = \frac{1}{N} \sum_{i=1}^N W_{i\tau}
\label{awpc}
\end{equation} 
for a system of $N$ individuals, whose wealth follows the rule of proportionate growth (\ref{prop_eff}),  can be approximated for large $N$ by $\overline{W}_\tau \approx E(W_\tau) = W_0 e^{\tau \Delta t(\mu + \sigma^2/2)}$. Expressing the wealth in units of the average wealth per capita 
\begin{equation}
w_{i\tau}  = \frac{W_{i\tau}}{\overline{W}_\tau}
\label{normalised}
\end{equation}
one gets a normalised wealth which has the mean $\overline{w}_\tau = 1$. The evolution of the normalised wealth can be approximated by $w_\tau = W_\tau/ E(W_\tau)$ which by construction has a log-normal distribution $\log\mathcal{N}(-\sigma^2 \tau/2 , \sigma^2 \tau)$ and the mean $E(w_\tau) = 1$. The dependence on $\mu$ disappears as it is cancelled by the normalisation.
The standard deviation of the normalised variable $w_\tau$ grows exponentially $\sqrt{e^{2 \tau \sigma^2} - 1} \sim e^{\tau \sigma^2}$  for large $\tau$. This means that the distribution gets quickly very broad and very skew. The median, $e^{-\sigma^2\tau/2}$, exponentially drops to zero as $\tau$ increases, meaning that the distribution rapidly gets concentrated at zero. This effect is even more clearly seen in the time dependence of quantiles of the distribution. For any $p<1$ the quantile is equal $q(p) = e^{\kappa_p \sigma \sqrt{\tau} - \sigma^2\tau/2}$ where $\kappa_p$ is the corresponding quantile for standardised normal distribution. For large $\tau$ the second term in the exponent is dominant and it makes the quantile $q(p)$ exponentially tend to zero as $\tau$ tends to infinity. In effect, for large $\tau$ almost one hundred percent of the distribution is concentrated at zero. At the same time the mean is equal one, this means that the distribution must develop a long tail towards infinity. Enormous inequalities arise in the system. Indeed, the Gini coefficient quickly approaches one: $G=\mathrm{erf}(\sigma \sqrt{\tau}/2) \approx 1 - \frac{2}{\sqrt{\pi}\sigma \sqrt{\tau}} e^{-\sigma^2 \tau/4}$ as $\tau$ grows. For large time, $\tau \gg \sigma^{-2}$, this extreme situation is realised by a highly uneven distribution of wealth with a single individual being tremendously rich and all remaining ones being extremely poor. The rich one amasses almost entire wealth of the whole system. Some regulatory mechanisms are needed to stabilise the system.

Let us make a few remarks. The stochastic process, Eq. (\ref{prop_eff}), describing the evolution of wealth can be modified in many ways. One can for example 
explicitly include earned income $E_\tau$, consumption $C_\tau$, inheritance, 
(donations, gifts and bequests) $I_\tau$, {\em etc.} into the evolution equation: $W_\tau = \Lambda_\tau W_{\tau-1} + E_\tau - C_\tau + I_\tau + \ldots$, {\em etc.} (see \cite{ds,bb} for review). One can also add taxes, interest rates for savings, {\em etc.} This would however make the modelling more complex and it would require additional assumptions about micro-foundations to simulate the behaviour of individuals and to describe mutual relations between all the stochastic processes $W_\tau$, $E_\tau$, $C_\tau$, $I_\tau$, {\em etc.} This would also drive the analysis towards a multi-parametric scenario. From the macro-economic perspective, however, such additive stochastic fluctuations in the macro-economic scale play a secondary role as compared to the multiplicative ones. They can of course enhance (or reduce) the rate of wealth inequality growth if they are positively (or negatively) correlated with the primary process of wealth growth. For example if the income $E_\tau$ is positively correlated with the wealth $W_\tau$ then rich ones quicker get richer and the wealth inequality grows faster. When it comes to modelling the same effect can be achieved, however, by correlating the multiplicative factors $\Lambda_\tau$ with the wealth $W_\tau$ in Eq. (\ref{prop_eff}). So in order to concentrate on dominant effects one can neglect secondary contributions and stay within the framework of purely multiplicative evolution which plays the primary role in shaping wealth distribution in the macro-economic scale.

\subsection{Kesten stochastic processes}

As mentioned, the law of proportionate effect in the form of Eq. (\ref{prop_eff}) leads to a non-stationary evolution. A simple modification of the multiplicative rule can, however, stabilise the system and make it evolve towards a stationary state. Assume, as before, that wealth is generated according to the law of proportionate effect, Eq. (\ref{prop_eff}), but if its value drops below a certain predefined threshold, $W_{min}$, the new value is rejected and $W_\tau$ stays on the previous level $W_{\tau-1}$ 
\begin{equation}
W_\tau = \left\{ \begin{array}{ll} \Lambda_\tau W_{\tau-1}  & \mathrm{if}\  \Lambda_\tau W_{\tau-1}  > W_{min} \\ 
                                   W_{\tau-1} & \mathrm{otherwise} \end{array} \right. .
\label{kesten}
\end{equation}
The second element of the construction is to assume that the random factors $\Lambda_\tau$ are contractive {\it i.e.} $E(\Lambda_\tau) < 1$. Such stochastic processes are called Kesten processes \cite{k}. They are known to have a stationary state described by a probability distribution with a power-law tail for asymptotically large values. What makes Kesten processes attractive in this context is that on the one hand they generate power-law tails which are observed in many empirical studies on wealth and income distributions. On the other hand the presence of the lower bound can be interpreted as a sort of existence limit. The contractive nature of multiplicative fluctuations is less obvious in the economic context but as we shall see below the effect can be naturally introduced by taxation, or more generally by regulatory measures and economic interventionism. It is worth mentioning that Kesten processes were formulated in mathematical literature without any reference to economy. In economy there are many specific models which in one way or another implement the contractive multiplicative fluctuations and the lower barrier, as for example Champernowne's model of wealth accumulation \cite{c} or a model of inter-generation wealth transfer by Wold and Whittle \cite{ww}. Kesten processes have been also  independently rediscovered in econophysics literature \cite{ls}. We want to stress that one can expect a broad class of multiplicative stochastic processes with contractive random factors and a lower barrier to have a stationary state represented by a distribution with a power-law tail for large asymptotic values. This is a quite general, robust and universal explanation of the presence of power-law tails in many empirical distributions for statistical systems. In contrast to preferential attachement that explains power-law tails in growing systems \cite{y,sim,ba} Kesten processes explain their presence in systems which do not grow.

\subsection{Agent-based modelling}

So far we have discussed evolution of wealth of a single individual. A qualitatively new picture is obtained by extending the description to the multi-agent framework where one explicitly takes into account mutual interactions of agents (individuals) \cite{a,szsl,bm}. A single evolution equation Eq. (\ref{prop_eff}) is replaced by a set of equations which can be schematically written as
\begin{equation}
W_{i\tau} = \Lambda_{i\tau} W_{i\tau-1} + \sum_{j} T_{ij,\tau} 
\label{abm}
\end{equation}
where indices $i,j$ run over the set of individuals $1,\ldots, N$. The first term in Eq. (\ref{abm}) follows Gibrat's law as before. What is new is the sum of interaction terms $T_{ij,\tau}$ which represent direct wealth transfers between agents $i,j$ in the period $\tau$. The interaction terms are antisymmetric $T_{ji,\tau} = - T_{ij,\tau}$, so they do not change the total wealth but only generate a wealth flow in the system.  The interactions can be chosen to be linear in wealth $T_{ij,\tau} = J_{ij} W_{j\tau-1} - J_{ji} W_{i\tau-1}$ with some wealth reallocation coefficients $J_{ij}$ controlling the intensity of wealth transfer. In this case the model is often called Bouchaud-M\'{e}zard model. The original model was formulated in the continuous-time limit \cite{bm}. If $J_{ij} = J_{ji}$ then the transfer is proportional to the wealth gradient $T_{ij,\tau} = J_{ij} \left(W_{j\tau-1} - W_{i\tau-1}\right)$. If reallocation constants are positive $J_{ij}>0$ the wealth gap between $i$ and $j$ decreases as a result of agent interactions, while for $J_{ij}<0$ it increases. Bouchaud, M\'ezard and followers considered interactions reducing wealth inequality. In this case the effective equations for wealth are contractive and the evolution equations belong to the class of Kesten processes, so one can expect the wealth distribution to have a power-law tail. Indeed, one can show that in this case, mean-field calculations lead to an inverse-gamma distribution in the statistical limit which has a power-law tail \cite{bm}. The power-law exponent depends on the ratio of the reallocation coefficient and the volatility of fluctuations in the Gibrat's growth factor. For sparse networks the limiting distribution is given by a generalised-inverse-gamma distribution \cite{mhs}. 
The system displays a highly non-trivial pattern of relaxation to the limiting distribution \cite{ls1,bpa,bss}. For finite size the system does not reach a stationary state. The non-stationarity is reflected in fluctuations of the scale parameter of the inverse-gamma distribution \cite{ls2}. 

The regime of the model that corresponds to aggressive economy, $J_{ij}<0$, favouring the rich-get-richer (and poor-get-poorer) interactions has not been studied yet. In this case the system is unstable \cite{bpa,bss} unless some regulatory, stabilising mechanism is introduced. 

\subsection{State interventionism}

Regulatory mechanisms are an important factor that determines the distribution of wealth. Policy makers can influence wealth distribution by changing legal regulations, taxes, investments, welfare, {\em etc.}. The problem is multidimensional, very complex and difficult to quantify so we again use Ocamm's razor to make it as simple as possible to focus on the most dominant impact of the public sector on wealth distribution of individuals. We assume  that each individual pays a linear wealth tax to the public sector and that the public money is then uniformly distributed among all individuals
\begin{equation}
W_{i\tau} = (1-\beta) W'_{i\tau} + \frac{\beta}{N} \sum_{j=1}^N  W'_{j\tau} \ ,
\label{capital_tax}
\end{equation}
where $W'_{i\tau} = \Lambda_{i\tau} W_{i\tau-1}$ {\em i.e.} individual changes follow the rule of proportionate growth, Eq. (\ref{prop_eff}). As we shall see, already this simple version of regulatory mechanism generates a non-trivial solution with a Pareto tail in wealth distribution. We have intentionally skipped direct interactions between agents in Eq. (\ref{capital_tax}) to focus on the effect of taxation and redistribution. The interactions, $T_{ij,\tau}$'s, will be restored later in the full model. The first term on the right hand side of Eq. (\ref{capital_tax}) is contractive. The second term, in turn, makes the minimal wealth never be smaller than $\beta$ times the current average per capita, so in a sense it introduces a lower bound on wealth. In other words the stochastic process, Eq. (\ref{capital_tax}), belongs to a broadly understood class of Kesten processes and thus one can expect it to have a stationary distribution with a Pareto tail. Below we show that this is indeed the case by solving analytically the model, Eq. (\ref{capital_tax}), for large $N$ in the continuous time limit. In this limit it is convenient to replace the tax rate $\beta$ by the corresponding continuously compounded tax rate $\beta_0$ which is related to $\beta$ as $1-\beta = e^{-\beta_0 \Delta t}$, where $\Delta t$ is the duration of the fiscal period. Clearly $\beta \approx \beta_0 \Delta t$ for $\beta_0 \Delta t \ll 1$. Under the assumption that the multiplicative fluctuations $\Lambda_{i\tau}$ are {\em i.i.d.} lognormal factors $\log\mathcal{N}(\mu_0 \Delta t, \sigma^2_0 \Delta t)$ the set of Eqs. (\ref{capital_tax}) reduces to a set of $N$ independent stochastic differential equations (for $N\rightarrow \infty$) for the normalised wealth defined in Eq. (\ref{normalised})
\begin{equation}
d w_i(t) = \beta_0 (1-w_i(t)) dt + \sigma_0 w_i(t) dB_i(t) \ . 
\label{sde} 
\end{equation}
Here we replaced $w_{i\tau}$ by the corresponding function $w_{i}(t)$ of a continuous time $t=\tau \Delta t$. The equations are independent of each other and they are identical for each $i$, so one can skip the index $i$. The corresponding Fokker-Planck equation reads
(see for instance \cite{vk})
\begin{equation}
\partial_t p(w,t) = \beta_0 \frac{\partial}{\partial w} \left((w-1)p(w,t)\right) + \frac{\sigma^2_0}{2} 
\frac{\partial^2}{\partial w^2} \left(w^2 p(w,t)\right) \ .
\label{evol}
\end{equation}
This equation describes evolution of the probability density function  $p(w,t)$
of wealth distribution at time $t$. Clearly the evolution preserves the mean $\int dw w p(w,t)=1$, since by construction the normalised wealth $w_i(t)=W_i(t)/\overline{W}(t)$ is expressed in terms of the average wealth $\overline{W}(t)$ at given $t$, and thus $\overline{w}(t)=1$. For any positive value of the tax rate $\beta_0>0$, Eq. (\ref{evol}) has a stationary state $\partial_t p_*=0$ which is reached $p(w,t) \rightarrow p_*(w)$ in the limit $t\rightarrow \infty$. One can find an explicit form of the probability density function for the stationary distribution by solving Eq. (\ref{evol}) with the left hand side replaced by zero. The solution reads
\begin{equation}
p_*(w) = \frac{c}{w^{1+\alpha}} e^{-\frac{\alpha-1}{w}}
\label{pstar}
\end{equation}
where $\alpha = 1 + 2\beta_0/\sigma^2_0 = 1 + 2\beta/\sigma^2$ and $c=(\alpha-1)^\alpha/\Gamma(\alpha)$. This is an inverse-gamma distribution. As expected, for large $w$ the distribution has a power-law tail. For any non-zero (positive) tax rate the exponent $\alpha$ is larger than one. One can check by inspection that the mean of the distribution (\ref{pstar}) is equal one. We see that the combination of taxation and redistribution generates power-law tails in the distribution of wealth. The effect is quite universal and has the same origin as in Kesten processes: taxation makes the process contractive while redistribution generates a lower bound. When the tax rate goes to zero ($\beta\rightarrow 0^+$) then $\alpha \rightarrow 1^+$. For $\alpha=1$ the function on the right hand side of Eq. (\ref{pstar}) is not integrable and the formula breaks down \cite{bm}. In this case the system is described by the stochastic process (\ref{prop_eff}) which never reaches a stationary state, as discussed.

It is worth mentioning that equations (\ref{sde}), (\ref{evol}) and (\ref{pstar}) are identical as equations for the mean-field version of the Bouchaud-M\'{e}zard model \cite{bm,ls,bpa} with the reallocation constant $J$ replaced by the tax rate $\beta_0$. The equations have a different origin, though. In the original Bouchaud-M\'{e}zard model \cite{bm} wealth inequality was reduced by agent interactions, whereas in the present version of the model (\ref{capital_tax}) it is reduced by taxes and redistribution which provide a stabilising factor for the system. Now we are going to define our model where we confront aggressive interactions between agents with macro-economic actions undertaken by policy makers to reduce wealth inequality in the global scale.

\section{The model}

\subsection{Wealth dynamics}

The main assumption behind the model is that the wealth distribution in a macro-economic system is driven by three processes: stochastic growth described by the law of proportionate effect -- Eq. (\ref{m1}), the flow of wealth between individuals -- Eq. (\ref{m2}), and a global mechanism representing economic interventionism of the public sector -- Eq. (\ref{m3})
\begin{align}
 W'_{i\tau}  = & \Lambda_{i\tau} W_{i\tau-1} \label{m1} \ , \\
 W''_{i\tau} = & W'_{i\tau} + \sum_{j=1}^{Q_i} T_{ij,\tau} \label{m2} \ , \\ 
 W_{i\tau}  = & (1 - \beta) W''_{i\tau} + \frac{\beta}{N} \sum_{j=1}^N W''_{j\tau} \ .
 \label{m3}
\end{align} 
The equations form an iterative system that describes a discrete-time evolution of wealth
$\{W_{1\tau},\ldots,W_{N,\tau}\}$ of $N$ individuals. 
The proportionate growth is encoded in the statistics of growth factors $\Lambda_{i\tau}$'s. In the simplest version of the model one can assume that $\Lambda_{i\tau}$'s are independent identically distributed ({\em i.i.d.}) random numbers.
The growth rates $r_{i\tau} \equiv \log \Lambda_{i\tau}$ are {\em i.i.d.} random numbers as well. For sake of simplicity it is also convenient to assume that the growth rates are Gaussian random variables $\mathcal{N}(\mu,\sigma^2)$ or equivalently that $\Lambda_{i\tau}$'s are {\em i.i.d.} lognormal random variables $\log \mathcal{N}(\mu,\sigma^2)$. The statistical properties of growth rates is controlled by the expected growth rate $\mu = E(r_{i\tau})$ and the volatility $\sigma^2=E((r_{i\tau}-\mu)^2)$ which is a measure of the magnitude of statistical fluctuations of growth rates around the trend $\mu$. 

Economic interactions between individuals can be represented as a network of interactions with nodes corresponding to individuals $i=1,\ldots, N$ and edges $ij$ to direct interactions. The node degree $Q_i$ is equal to the number of individuals with whom an individual $i$ directly interacts. The interaction terms are antisymmetric $T_{ij,\tau} = -T_{ji,\tau}$, because the wealth flow from $i$ and $j$ is opposite to that from $j$ to $i$. The flow preserves the total wealth in the system as one can see summing both sides in Eq. (\ref{m2}) over $i$: $\sum_{i=1}^N W''_{i\tau} = \sum_{i=1}^N W'_{i\tau}$. The total wealth is preserved also in Eq. (\ref{m3}). Wealth is generated only by the growth factors in Eq. (\ref{m1}). In the large $N$ limit, the expected average wealth per capita increases exponentially $E(W_\tau) = e^{\tau \Delta t (\mu+\sigma^2/2)} W_0$. This follows from the independence of $\Lambda$'s and $W$'s which  for large $N$ allows one to factorise the expected value in Eq. (\ref{m1}):  $E(\overline{W}_\tau) =  E(\overline{\Lambda}_\tau) E(\overline{W}_{\tau-1}) =
e^{\Delta t (\mu + \sigma^2/2)} E(\overline{W}_{\tau-1})$.

It is convenient to assume the wealth transfer $T_{ij,\tau}$ to be linear in wealth in the sense that they rescale as $T_{ij,\tau} \rightarrow s T_{ij,\tau}$ under rescaling of  $W_{i\tau} \rightarrow s W_{i\tau}$ by the same factor $s$ for all $i$. Under this assumption all three Eqs. (\ref{m1})--(\ref{m3}) are invariant under the change of scale. This means in particular that they are invariant with respect to the choice of monetary units (currency) the wealth is expressed in. It is convenient to use the current average wealth per capita Eq. (\ref{awpc}) as a unit of wealth. Expressed in these units the wealth values at time $\tau$ correspond to the normalised values $w_{i\tau}$, Eq. (\ref{normalised}), so the average wealth per capita is $\overline{w}_\tau=1$. The normalised values (are dimensionless) and can be easily compared to one another. One has to remember however that the average wealth per capita, which serves as a unit of wealth, varies in time. Its expected value changes exponentially $E(\overline{W}_\tau) = e^{\tau \Delta t (\mu +  \sigma^2/2)} \overline{W}_0$.

The simplest linear formula for the wealth transfer is $T_{ij,\tau} = J_{ij} W'_{j\tau} - J_{ji} W'_{i\tau}$ with some constant parameters $J_{ij}$ \cite{bm}. These parameters control the intensity and the direction of wealth flow between individuals. 
The total wealth outflow from agent $i$ at time $\tau$ is proportional to his current wealth $W'_{i\tau}$ and to an outflow coefficient, $J_i$, being a sum of intensity factors $J_i=\sum_j J_{ji}$. The complement of the outflow coefficient, $S_i = 1-J_i$, is a fraction of wealth kept by agent $i$ as a sort of inactive capital, {\em e.g.} real estate or savings. In the symmetric case $J_{ij}=J_{ji}$ the flow equation (\ref{m2}) takes the form $W''_{i\tau} =  W'_{i\tau} + \sum_{j=1}^{Q_i} J_{ij} (W'_{j\tau} - W'_{i\tau})$ where the primed values denote wealth of $i$ before the wealth transfer and the double primed ones - after it. As follows from this equation the capital is transferred from the richer individual to the poorer one if $J_{ij}>0$ and in the opposite direction if $J_{ij}<0$. This means that the wealth gap between $i$ and $j$ gets smaller for positive $J_{ij}$'s, and larger for negative $J_{ij}$'s . Negative $J_{ij}$'s provide a microscopic implementation of the Mathew effect realising the rich-gets-richer (and poor-get-poorer) scenario. In this version of the model wealth is not bounded from below and can become negative. Once it becomes negative it gets later even more negative as a result of the poor-get-poorer feedback. One can eliminate this effect by introducing a limit on the maximal loss of wealth in a single fiscal period. We assume that the maximal wealth flow from $i$ to $j$ is proportional to the current wealth of $i$. The proportionality coefficient we denote by $\gamma_i$. This leads to the following expression for the wealth transfer $T_{ij,\tau} = \min\left(-\gamma_{i} W'_{i\tau},\max\left(\gamma_j W'_{j\tau},J_{ij} W'_{j\tau} - J_{ji} W'_{i\tau}\right)\right)$. The kernel of this expression, $J_{ij} W'_{j\tau} - J_{ji} W'_{i\tau}$, is exactly the same as in the Bouchaud-M\'{e}zard model \cite{bm} but here it is bounded by the maximal loss limit, $-\gamma_{i} W'_{i\tau}$, and the maximal gain limit, $\gamma_j W'_{j\tau}$. The two limits are related to each other since what is a gain of $i$ is a loss of $j$ and vice versa. The expression is antisymmetric $T_{ij,\tau}=-T_{ji,\tau}$, as it should be. Because agent $i$ interacts with $Q_i$ agents, agent $i$ may in the worst case loose $Q_i \gamma_i W'_{i\tau}$. The coefficient $g_i=Q_i \gamma_i$ in this expression is the fraction of the capital of agent $i$ which may be maximally lost in one fiscal period. If $g_i<1$, the individual $i$ may loose a part of his or her wealth, if $g_i=1$ he may loose everything, and if $g_i>1$ he may loose more than he has and in effect he may end up in depts. The parameter $g_i$ reflects the risk attitude of agent $i$. For a risk-avoiding individual: $g_i<1$, for a risk-seeking one: $g_i>1$, and for a risk-neutral one: $g_i=1$. Replacing $\gamma_i$'s by the risk attitude coefficients $g_i$'s in $T_{ij,\tau}$ we get 
\begin{equation}
T_{ij,\tau} = \min\left(-\frac{g_i}{Q_i} W'_{i\tau}, \max\left(\frac{g_j}{Q_j} W'_{j\tau},
J_{ij} W'_{j\tau} - J_{ji} W'_{i\tau}\right)\right) \ .
\label{wealth_exchange}
\end{equation}
This expression is used in Eq. (\ref{m2}) to model wealth flow in the system. 

Eq. (\ref{m3}) models the influence of the public sector on evolution of the wealth distribution. This equation can be modified in many ways. For example one can replace the linear tax by a progressive tax by making the coefficient $\beta$ depend on wealth. One can introduce other taxes as for instance income tax which is based on the income $W_{i\tau}-W_{i\tau-1}$ rather than on the wealth $W_{i\tau}$ itself. One can also replace the uniform redistribution by a targeted redistribution depending on the wealth level of individuals. The model has a flexible structure which can be adapted to study various effects.

\subsection{Mean-field approximation}

In this section we consider a mean-field approximation of the model. The idea is to assume that all individuals are sort of typical. They interact with each other ($Q_i = N-1$) with the same average intensity $J_{ij}=J/(N-1)$. They have the same risk attitude $g_i=g$. All wealth outflow coefficients are identical $J_i = J$. It is natural to assume that $J<1$. The wealth transfer expression, Eq. (\ref{wealth_exchange}), simplifies to
\begin{equation}
T_{ij,\tau} = \frac{1}{N-1}\min\left(-g W'_{i\tau}, \max\left(g W'_{j\tau},
J( W'_{j\tau} - W'_{i\tau}) \right)\right) \ .
\label{tij_mf}
\end{equation}
\begin{figure}
\includegraphics[width=0.5\textwidth]{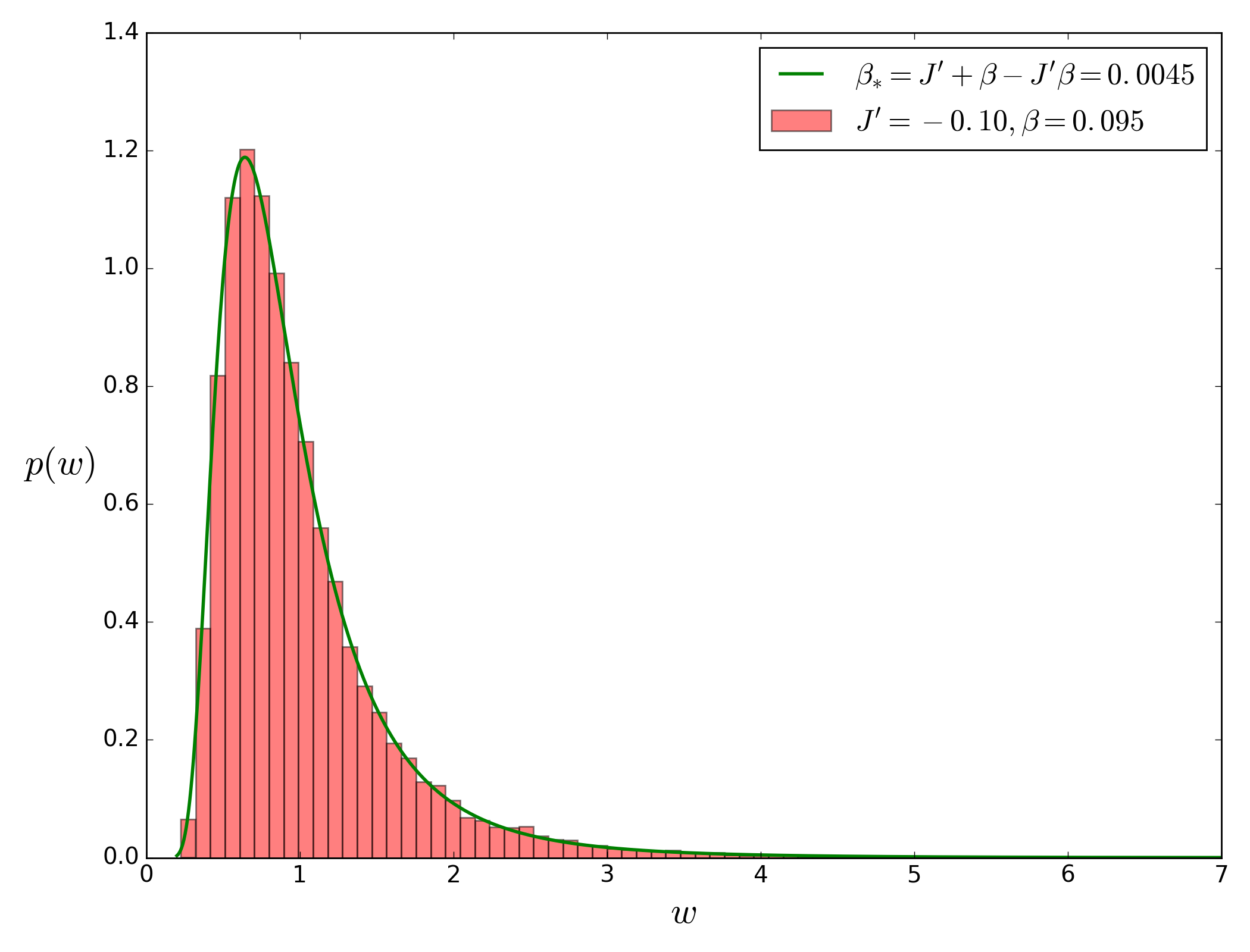}
\caption{Comparison of the distribution of wealth for the stationary state obtained in Monte-Carlo simulations of the model with the theoretical prediction, Eq. (\ref{pstar}) for $\alpha=4.6$. The parameters used in the simulations are $N=1000$, $J'=-0.1$, $\beta=0.095$, $\sigma=0.05$ and $g=1$. This yields $\beta_* = \beta + J' - \beta J' = 0.0045$ and $\alpha = 1 + 2\beta_*/\sigma^2=4.6$, Eq. (\ref{pstar}). The histogram was collected from one hundred samples. 
\label{fig_distr}}
\end{figure}
\begin{figure}
\includegraphics[width=0.5\textwidth]{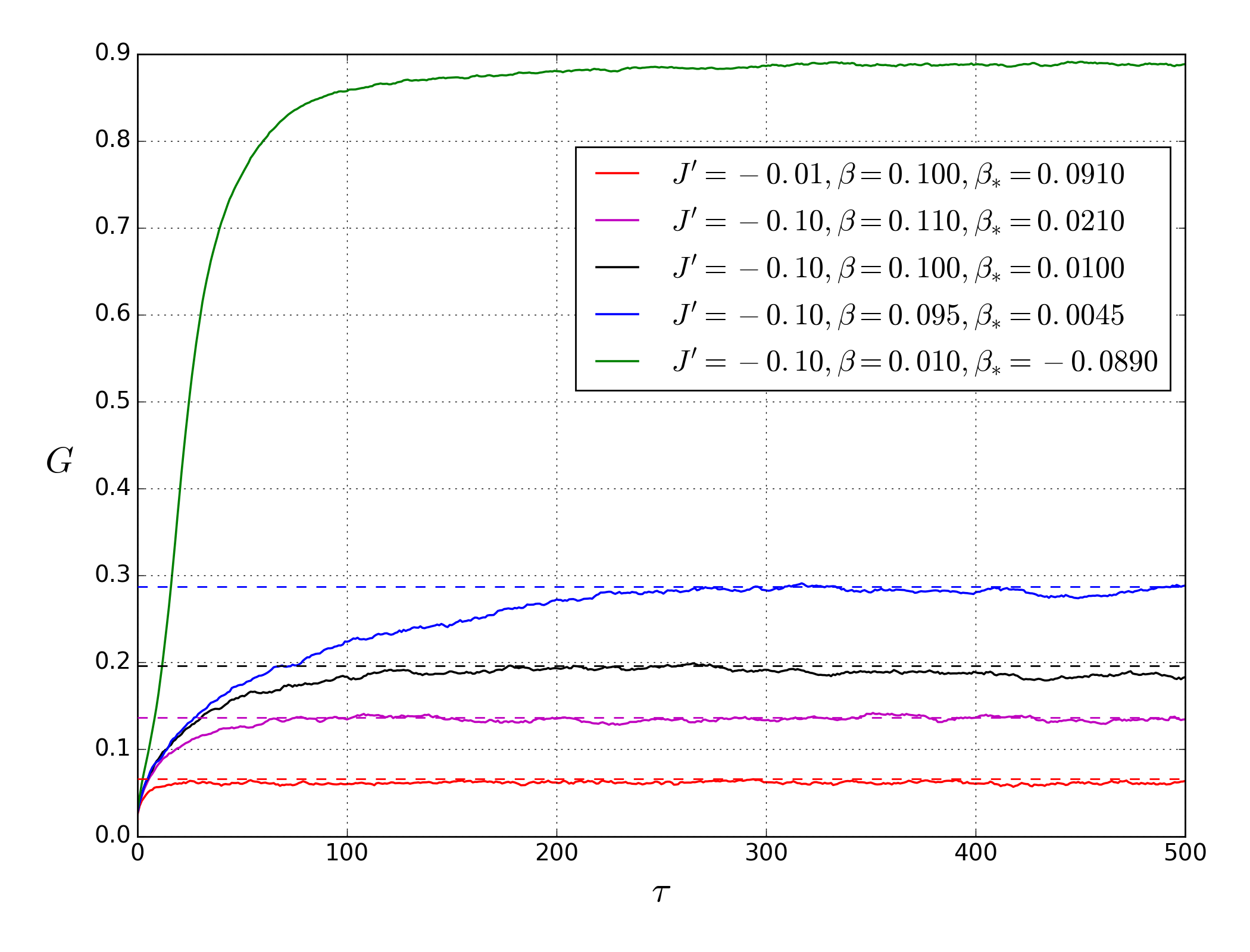}
\caption{Time evolution of the Gini index obtained in Monte-Carlo simulations of the model for different parameters (see the inlet) which correspond to $\beta_*=\{0.091,0.021,0.01,0.0045,-0.089\}$ (from bottom to top). Other parameters used in the simulations are $N=1000$, $\sigma=0.05$ and $g=1$. The horizontal lines show theoretically predicted limiting values of the Gini index $G=\Gamma(2\alpha-1)/\Gamma(\alpha) \left\{ _2F_1(\alpha-1,2\alpha-1;\alpha;-1)/\Gamma(\alpha) + (1-\alpha) 
_2F_1(\alpha,2\alpha-1;\alpha+1;-1)/\Gamma(\alpha+1)\right\}$
for the stationary distribution, Eq. (\ref{pstar}), with $\alpha=1+2\beta_*/\sigma^2$ for $\beta_*>0$. The prediction breaks down for negative $\beta_*$ (the uppermost curve in the figure). In this case the corresponding wealth distribution has a multimodal structure as discussed in the next section. The symbol $_2F_1(a,b;c;z)$ stands for the hypergeometric function \cite{gr}. \label{fig_distr_gini}}
\end{figure}
For $0<J<1$ and for risk-neutral attitude, $g=1$, one can further
simplify the last equation  by skipping the min-max bounds which are automatically fulfilled in this case. This gives $T_{ij,\tau} = \frac{J}{N-1}( W'_{j\tau} - W'_{i\tau})$. In effect, Eq. (\ref{m2}) takes a simple form $W''_{i,\tau} = (1-J') W'_{i,\tau} + \frac{J'}{N}\sum_{j=1}^N W'_{j,\tau}$ where $J'=J\frac{N}{N-1}$. For large $N$: $J'\approx J$. This equation is identical as Eq. (\ref{m3}) with $\beta$ replaced by $J'$. Moreover inserting $W''_{i\tau}$ from the last equation to Eq. (\ref{m3}) one can reduce Eq. (\ref{m2}) and Eq. (\ref{m3}) to a single equation 
\begin{equation}
W_{i\tau}  = (1 - \beta_*) W'_{i\tau} + \frac{\beta_*}{N} \sum_{j=1}^N W'_{j\tau} 
\label{effective}
\end{equation}
which is again of the same form as Eq. (\ref{m3}) with
\begin{equation}
\beta_* = \beta + J' - \beta J' \ .
\label{beta_star}
\end{equation}
Eq. (\ref{effective}) is identical as Eq. (\ref{capital_tax}) with $\beta$ replaced by $\beta_*$ and thus it has the same solution. In particular, in the continuous time limit and for large $N$, the probability density function of the stationary state is given by Eq. (\ref{pstar}) with $\alpha = 1 + 2\beta_*/\sigma^2$. This solution holds as long as $J > 0$. We have checked numerically that the limiting distribution given by Eq. (\ref{pstar}) is also a very good approximation for large but finite systems, for $N$ in the order of a hundred or larger, and for discrete time evolution (\ref{m1})-(\ref{m3}), as long as the parameters $\sigma$ and $\beta_*$ are significantly smaller than one. This is illustrated in Fig. \ref{fig_distr} and Fig. \ref{fig_distr_gini} where we compare results of Monte-Carlo simulations with theoretical predictions following Eq. (\ref{pstar}). The role of the parameter $\beta_*$ can be easily understood when one rewrites Eq. (\ref{effective}) in the following way: $\Delta W'_{i\tau}  =  -\beta_* \left(W'_{i\tau} - \overline{W'}_\tau\right)$
where  $\overline{W'}_\tau = \frac{1}{N} \sum_{j=1}^N W'_{j\tau}$ is the average wealth per capita and $\Delta W'_{i\tau} = W_{i\tau}-W'_{i\tau}$ is a change of wealth of agent $i$ caused by the effective wealth flow in the system. For $\beta_*>0$ the right hand side of this equation describes an attractive force which tries to keep the wealth values close to the mean value and to reduce wealth inequality in the system. This force results from macro-economic interactions, taxation and redistribution. It counteracts the wealth differentiation coming from multiplicative fluctuations generated by the law of proportionate effect, Eq. (\ref{m1}). The situation changes for $\beta_*<0$ in which case the wealth values are repeled from the mean value. In this case wealth inequality is strengthened and the system undergoes a non-stationary evolution. Empirical data suggest that the world macro-economy has entered a period of increasing wealth inequality in the global scale \cite{p,sz,p1}. We study this regime in detail in the next section.

\subsection{Rich-get-richer macroeconomy}

The Matthew effect of accumulated advantage is modelled by negative $J_{ij}$'s imitating macro-economic forces which increase wealth inequality. The process can be slowed down by risk-avoiding behaviour of individuals and/or regulatory measures including taxation, social security, welfare, {\em etc.}. The evolution of the system is shaped by all these factors. In the mean-field approximation for $J<0$, the wealth flux, Eq. (\ref{tij_mf}), is negative: $T_{ij,\tau}<0$ if individual $i$ is poorer than $j$. This means that wealth flows from $i$ to $j$ and thus $i$ gets poorer and $j$ richer. The wealth gap between $i$ and $j$ increases. In the worst case individual $i$ who owns wealth $W'_{i\tau}$ may loose in total $g W'_{i\tau}$ in one fiscal period and thus his wealth may drop to $(1-g) W'_{i\tau}$ as a result of interactions with other agents. For the risk neutral attitude, $g=1$, the wealth, $W'_{i\tau}$, may maximally drop to zero. In effect in this case there are no individuals with a negative wealth in the system. This would not be the case if there were risk-seeking individuals, $g>1$. 

Before presenting results of simulations of the model for $J<0$ let us collect in one place all significant parameters and describe their role in the underlying macro-economic system. The dynamics of wealth distribution is mainly governed by the parameter $|J|$ which reflects the strength of the positive feedback coupling in the rich-gets-richer mechanism. The larger $|J|$ the larger is the growth rate of the wealth gap. The parameter $\beta$ imitates the strength of regulatory mechanisms applied by policy makers to prevent wealth inequality from growing. In this version of the model, the system is regulated by a linear capital tax combined with the uniform redistribution of tax revenue, and $\beta$ is just the tax rate. The volatility $\sigma$ is a measure of fluctuations of multiplicative factors in the law of proportionate effect. It reflects the scale of fluctuations of return rates on capital around the main trend. Larger $\sigma$'s correspond to better conditions for risky enterprises and to larger fluctuations of wealth. If $\sigma$ is large, poor individuals have a large chance to become rich and vice versa. We assume in our considerations the risk attitude of individuals to be either risk-avoiding $g<1$ or risk-neutral $g=1$. This ensures that the values $W_{i\tau}$ are non-negative for all individuals during the whole evolution. 

We have simulated systems of size $N$ ranging from one hundred to ten thousand. We have noted however that already for $N$ in the order of a hundred the behaviour of the system is qualitatively the same as for larger $N$. All figures in this paper are for $N=1000$. The
simulations are performed by iterating Eqs. (\ref{m1})--(\ref{m3}) with $Q_i=N-1$ and $T_{ij,\tau}$'s given by Eq. (\ref{tij_mf}). The stochastic component of the model is encoded in the multiplicative factors $\Lambda_{i\tau}$ in Gibrat's law, Eq. (\ref{m1}), which are generated as {\em i.i.d.} log-normal random numbers $\log \mathcal{N}(\mu,\sigma^2)$. All measurements are done on normalised values of wealth, Eq. (\ref{normalised}), which are by construction independent of $\mu$. In order to improve numerical stability of the simulations and to reduce exponential effects we set $\mu=-\sigma^2/2$ and additionally substitute wealth values $W_{i\tau}$ by the normalised values, Eq. (\ref{normalised}), after each iteration of Eqs. (\ref{m1})--(\ref{m3}) for all agents.  
\begin{figure}
\includegraphics[width=0.7\textwidth]{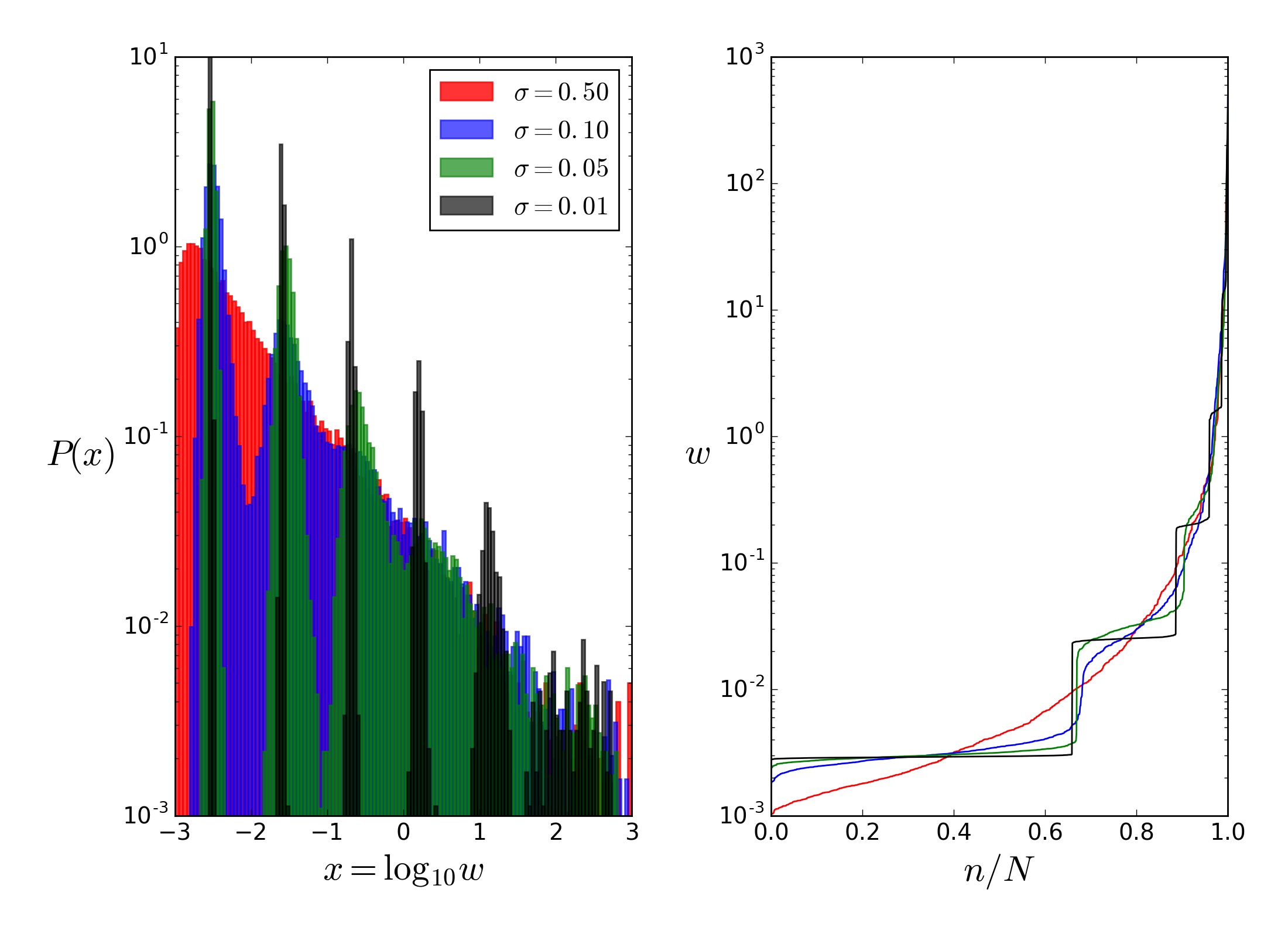}
\caption{Dependence of wealth distribution on $\sigma$. The left chart shows histograms of wealth distribution for the simulations of the system with $\sigma=\{0.01,0.05,0.1,0.5\}$ and $N=1000,\beta=0.001, g=1,J=-0.3$ after $\tau=1000$ steps of evolution. Each histogram was collected from one hundred samples. For $\sigma=0.01$ the histogram consists of peaks which are narrow and separated from each other. They get broader when $\sigma$ is increased. They overlap and form a single broad distribution for $\sigma=0.5$. The right chart shows corresponding data (for one sample) represented as a plot of logarithm of wealth vs. normalised rank. The normalised rank $n/N$ is ordered from the poorest individual, $n=1$, to the richest one, $n=N$. For intermediate values of $\sigma$ (e.g. $\sigma=0.1$) one can see a smeared staircase structure indicating a tendency of the population to decay into economic classes.
\label{fig_sigma1}}
\end{figure}

For negative reallocation coefficients such that $\beta_* \le 0$ (\ref{beta_star}) interactions increase wealth inequality. If there were no regulatory mechanisms the wealth distribution would never reach a stationary state. A stationary state is reached thanks to taxation and redistribution, Eq. (\ref{m3}), which play the role of a stabilising factor in the macro-economic scale. The stationary wealth distribution is however highly non-trivial. It depends on many factors, mostly on the economic volatility $\sigma$. In Fig. \ref{fig_sigma1} we show histograms representing wealth distributions of stationary states reached in Monte-Carlo simulations of the system with different values of $\sigma$. We see that for large $\sigma$ the histogram is unimodal. When $\sigma$ gets smaller the histogram becomes multimodal. A quantitative analysis of this effect is presented in Appendix. Here we describe it in a qualitative way. Imagine that one takes a snapshot of the population at some moment and orders individuals according to their wealth $w_1 < w_2 < \ldots < w_N$. If there were no stochastic component in the model ($\sigma=0$) the wealth order (rank) would be preserved during the evolution. An agent $i$ which is poorer than $j$ at the beginning, $w_i<w_j$, would remain poorer during the whole evolution. The wealth inequality would increase. The situation is very similar to a Coulomb gas of repelling particles in one dimension. If the gas is left to itself particles repel and move away from each other to infinity. The gas never reaches a stationary state. The system can be stabilised however by external forces or external mechanisms. For example an external confining potential or a boundary may keep particles in a confined region. In this case particles cannot escape to infinity and the system can equilibrate after some time. The equilibrium state results from the competition of repulsion between particles and the confinement coming from the external potential or boundary. The evolution preserves order of particle positions $x_1< x_2< \ldots <x_N$. Assume that we add a stochastic component to the evolution which makes particles randomly change their positions from time to time: $x_i \rightarrow x_i + \Delta x_i$ by $\Delta x_i$ being a random number of mean zero and variance $\sigma^2$. If $\sigma$ is much smaller than the spacing between particles $|x_i - x_{i-1}|$ the evolution will preserve the order $x_1< x_2< \ldots <x_N$.
However if $\sigma$ is comparable to the spacing, the order of particles will be reshuffled every now and then. Finally if $\sigma$ is much larger than the spacing, the evolution will be dominated by the stochastic component. Particle trajectories will form a laminar pattern for small $\sigma$ and a turbulent one for large $\sigma$ (see Appendix for details). The evolution of wealth in our model is very similar to that described above for the Coulomb gas in one-dimension. There are of course some differences: wealth changes are multiplicative rather than additive and the stabilising factor comes from economic regulations and not from a confining potential. The way the tax and wealth redistribution are implemented has some consequences for the wealth evolution. For small $\sigma$ the population bifurcates into a group of poorest individuals and the rest. The poorest group is driven towards the minimal wealth limit. The wealth level of this class gets separated from the rest of the population by a wealth gap. If the volatility is smaller than the wealth gap, individuals from the poorest class are not able to migrate to higher wealth classes and the poorest economic class stays separated for the rest of the evolution. Then the scenario may repeat in the remaining part of the population which may split again into a poorer class and the rest, etc. During the evolution there is a sequence of bifurcation points where new wealth gaps between economic classes arise, so the population decays into wealth classes. The detailed mechanism of wealth gap formation is given in the Appendix. In effect the distribution is multimodal. Modes of the distribution correspond to the most probable wealth values in economic classes. If the volatility is comparable to the wealth gaps the individuals may migrate between classes. The multimodal structures gets smeared in this case. This effect is clearly seen in Fig. \ref{fig_sigma1}.

Small fluctuations (small $\sigma$) are characteristic of over-regulated macro-economic systems. In a macroeconomic system which favours innovations, ideas, risky enterprises and economic freedom the scale of fluctuations is large. Risky ideas and risky investments may lead to a profitable business which can easily multiply wealth. On the other hand they may lead to a complete failure and to a financial disaster. Economic freedom is modelled by large fluctuations (large $\sigma$) of the multiplicative factors in the Gibrat law of proportionate effect, Eq. (\ref{m1}). 

\begin{figure}
\includegraphics[width=0.5\textwidth]{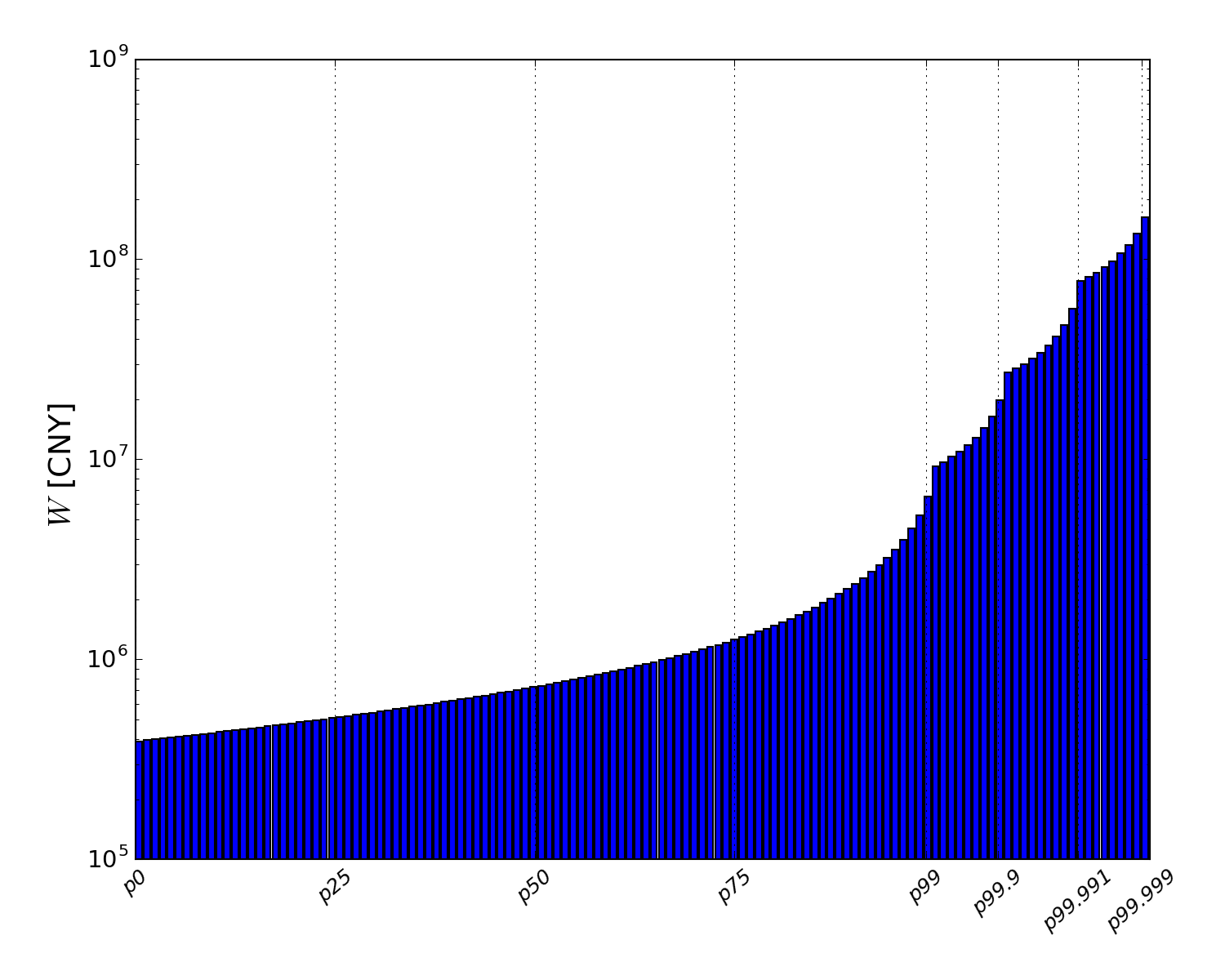}
\caption{The average wealth over a given percentile for the population in urban areas of China. The scale on the horizontal axis is linear up to $p=99$ while it is stretched in the last percentile in a nonlinear way to bring the fine structure within the richest part of the population into sharper focus. One can see characteristic steps which are predicted by the theory for systems with a small volatility (compare Fig. \ref{fig_sigma1}). The vertical axis shows net personal wealth calculated as a sum of personal non-financial assets and personal financial assets minus personal debt, in Chinese Yuan. The data source: http://wid.world/data  \cite{pyz} .\label{fig_china_urban}}
\end{figure}

To summarise, the model predicts economic stratification which manifests as a decay of the population into economic classes of different wealth levels. The wealth levels are separated by orders of magnitude (see Appendix). If the macroeconomic system is over-regulated the wealth segregation may be very strong. When economic freedom increases the economic volatility increases too, the segregation fades away and the division into separated classes disappears. For large fluctuations the stratification disappears completely. This phase is also characterised by fluctuations on the wealth rank list: someone being at the bottom may end up after some time on the top of the list and vice versa. Most of empirical data exhibit no clear division into separated wealth levels. But there are examples where one can see signs of economic stratification. An example is shown in Fig. \ref{fig_china_urban} which shows an empirical data on distribution of wealth for urban areas of China. The pattern fits to the picture predicted by the model, compare Fig. \ref{fig_sigma1}.

\begin{figure}
\includegraphics[width=0.7\textwidth]{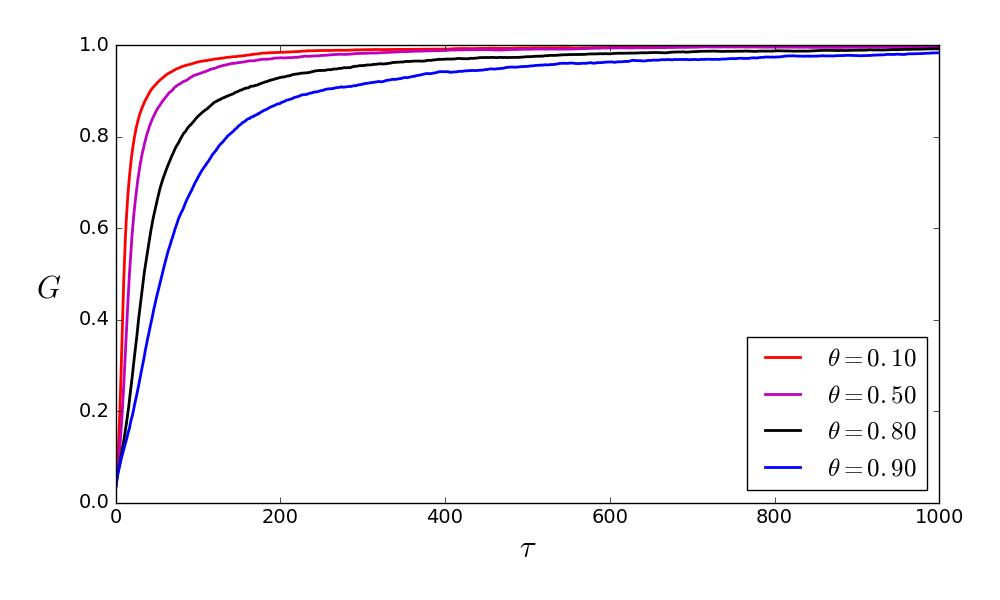}
\caption{Evolution of the Gini index for the system with different income tax rates $\theta$. Capital tax is switched-off, {\em i.e.} $\beta=0$. Other parameters used in the simulations are $N=1000$, $J=-0.3$, $\sigma=0.05$. For all $\theta$ the Gini index $G$ tends to one as $\tau$ increases. The system evolves towards the state of maximal inequality even if a large income tax rate is applied. Increasing income tax slows down the process of inequality growth but not the final inequality level which is given by $G=1$. \label{fig_gini_comp}}
\end{figure}

So far we have discussed dynamics of wealth distribution in the presence of wealth tax.
As a final remark we want to mention that the behaviour of the system dramatically changes if one replaces wealth tax by income tax. Wealth classes are not formed in this case anymore. Moreover, whatever the level of the income tax rate is the system drifts towards the state of maximal inequality with a single individual being much richer than the whole remaining population. The Gini index approaches one as one can see in Fig. \ref{fig_gini_comp}. What depends on the tax rate is the rate at which the state of maximal inequality is being approached. This situation significantly differs from that for wealth tax where one can reduce the value of the Gini index by increasing the tax rate, see Fig. \ref{fig_distr_gini}.

\section{Conclusions}

We have studied large scale emergent phenomena in a macro-economic system using agent-based modelling, population dynamics and stochastic evolution equations. Depending on whether macro-economic forces decrease or increase wealth gap between individuals the system undergoes distinct types of evolution. In the former case, the forces equilibrate the system and make the wealth distribution reach a stationary state described by the Pareto law. The exponent of the power-law (see Eq. (\ref{pstar})) depends on how strong are the stabilising forces in comparison to the macro-economic volatility. Generically if these forces are strong the wealth distribution is concentrated and there is no significant wealth inequality in the system. Only if they become weak as compared to the volatility of return rates, the system may develop a heavy tail with the Pareto exponent close to one which signals the occurrence of rich outliers in the population. When the stabilising forces completely disappear the system is not able to equilibrate. The system may be stabilised only by applying regulatory mechanisms and an economic intervention. The system becomes even more unstable if macroeconomic conditions favour aggressive capitalistic relations dominated by the rich-get-richer and poor-get-poorer feedbacks which disequilibrate the system. Left to itself the system will quickly develop the state of maximal inequality with the Gini index equal one. One can try to reduce the effect by applying regulatory measures, taxation and redistribution. The effect of introducing income tax and wealth tax have however different impacts on the system. Income tax slows down the process of rising inequalities but does not inhibit it. The system sooner or later reaches the state of maximal inequality with the Gini index equal one. The situation changes when wealth tax is imposed on individuals. The Gini coefficient of the final state can be decreased when the tax level is increased. Increasing wealth tax can apparently reduce wealth inequality in the system.

Depending on the scale of the volatility, the theory predicts different types of evolution of wealth distribution in the rich-get-richer (poor-get-poorer) macro-economy.  The volatility reflects the scale of fluctuations of return rates and the degree of economic freedom in the system. In an over-regulated macroeconomy the volatility is small, there is no room for risky investments. In a liberal macro-economy it is large. In a system with a very small volatility the evolution of wealth distribution exhibits a laminar pattern and leads to a formation of economic classes consisting of individuals on a similar wealth level. The classes are separated from each other by wealth gaps. After the class structure gets crystallised it is difficult for an individual to change the wealth level. The positions on the wealth rank list are frozen: rich individuals stay rich, poor ones stay poor. A wealth rank reshuffling is possible only within a wealth class. On the contrary, in a system with a very large volatility, the evolution exhibits a turbulent pattern which means that a poor individual may quickly become rich, and a rich one - poor. In this case positions on the wealth rank list are continuously reshuffled. For intermediate values of the volatility the two effects may coexist in different proportions reflecting the degree of economic freedom. One can expect to observe traces of the class structure in empirical data for macro-economic systems which have undergone a period of limited economic freedom recently (see Fig. \ref{fig_china_urban}).

The main idea behind the theoretical approach developed in this work was to implement the principle of accumulated advantage into the population dynamics describing interactions of individuals in a macro-economic system and to study its implications and structural consequences for the macroeconomic system, wealth inequality, or more generally, for wealth distribution. The focus of the study was on new emergent phenomena which can occur in evolution of macro-economic systems in the presence of disequilibrating effects which increase inequalities. We concentrated on the most dominant effects like economic stratification or the impact of taxation and redistribution on inequality dynamics. In the future one can extend the study and concentrate on specific issues. In particular one can go beyond the mean-field regime by considering a distribution of the wealth flow parameters, $J_{ij}$'s, some of which can be chosen positive and some negative. Similarly one consider a distribution of risk attitudes coefficients, $g_i$'s, to include risk-seeking individuals in the system. In this case there will be households with negative wealth that is with debts exceeding the total value of their assets. The spectrum of open problems is broad: one can also model collective phenomena \cite{cb}, the dynamics of the public sector and its interactions with the private sector. One can address the issue of optimal taxation and optimal economic growth \cite{b} or investigate the effect of changing macro-economic constraints on wealth distribution and stability of the system \cite{bjjknpz,b1}. 

\section{Appendix}

\begin{figure}
\includegraphics[width=0.5\textwidth]{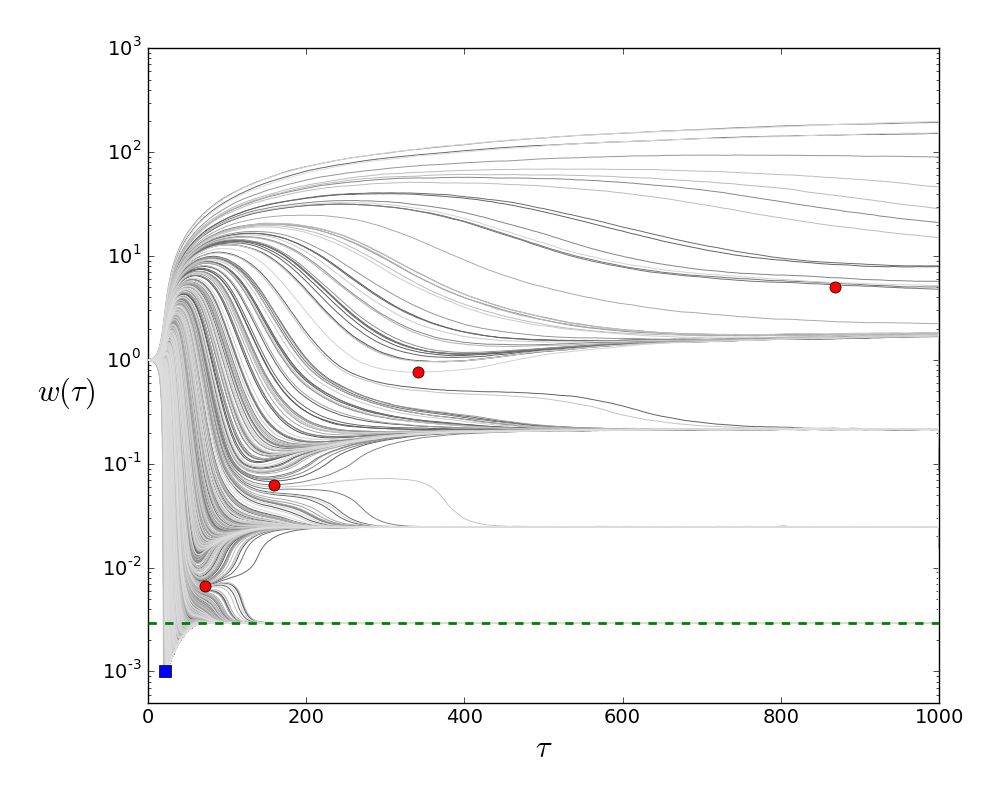}
\caption{Evolution of wealth distribution in a system with small growth rate fluctuations, $\sigma=0.001$. Remaining parameters used in the simulation are $N=1000$, $\beta=0.001$, $J=-0.3$ and $g=1$. Each line in the plot represents evolution of wealth of a single agent in logarithmic scale. The population decays into wealth classes in a series of bifurcations. The bifurcations have a characteristic pattern: the lower envelop of the upper wealth class descents to a local minimum and then ascents in a process of relaxation to its final asymptotic level while the upper envelop of the lower class descents. The bifurcation points are marked by red circles. The wealth gaps between consecutive wealth classes are minimal at the bifurcation points and they increase afterwards. The position of the blue square marks a theoretically calculated level of the absolute minimum,  Eq. (\ref{abs_min}). It agrees with the absolute minimum reached in the simulation. 
The dashed line shows the theoretically determined value of the asymptotic wealth level for the poorest class. The value depends on the number $N_1$ of individuals in this class. For the simulation in the figure $N_1=660$, which gives $w_{min}=0.0029$, Eq. (\ref{min_n1}) Again we see a good agreement between the simulation and the prediction. 
\label{fig_evolution}}
\end{figure}

In the appendix we discuss main features of the rich-gets-richer mean-field dynamics for the evolution given by Eqs. (\ref{m1}-\ref{m3}).  We begin by discussing evolution in a macro-economic system with small fluctuations, {\em i.e.} with a small volatility $\sigma$. First we choose a very small value of $\sigma$ to imitate an extreme situation. It will later provide us with a reference point for the discussion of evolution for larger $\sigma$'s - for more realistic situations. A typical pattern of evolution in the regime of very small volatility is shown in Fig. \ref{fig_evolution}.
Each curve in the figure represents evolution of wealth
of a single individual. The curves form a bundle whose shape reflects the process of wealth differentiation in the system. For $t=0$ all individuals have the same wealth $w_{i0}=1$ and the bundle is pinned to a single point. In the course of time the bundle decays into smaller bundles. The lower envelope of the bundle has a characteristic shape. It dives down to a point where it reaches the lowest level and then it bounces and lifts up to an asymptotic level. The wealth at the minimum can be determined as follows. Denote the normalised wealth of the poorest individual at time $\tau$ by $w_{min,\tau}$. If $w_{min,\tau}$ is much smaller than the wealth of other individuals then the effect of applying three equations, Eq. (\ref{m1})--(\ref{m3}) reduces to
\begin{equation}
w_{min,\tau} = (1-\beta) (1-g) \lambda_\tau w_{min,\tau-1} + \beta  \ ,
\end{equation}
where $1-\beta$ is the fraction of the wealth left after taxation,
$1-g$ is the maximal loss factor coming from the flow equation and
$\lambda_\tau$ is a lognormal factor $\log\mathcal{N}(-\sigma^2/2,\sigma^2)$ with $\mu=-\sigma^2/2$ which ensures the normalisation $E(\lambda_\tau)=1$. 
The additive term $\beta$ comes from the uniform redistribution, Eq. (\ref{m3}). For small $\sigma$ fluctuations can be neglected and $\lambda_\tau$'s can be substituted by the mean $E(\lambda_\tau)=1$. In this approximation the absolute minimum of the envelope can be estimated from the equation 
\begin{equation}
w_{min} = (1-\beta) (1-g) w_{min} + \beta  
\label{fp}
\end{equation}
which gives
\begin{equation}
w_{min} = \frac{\beta}{\beta + g(1 -\beta)} \ .
\label{abs_min}
\end{equation}
In particular for the risk neutral case, $g=1$, the last equation yields $w_{min}=\beta$. 
This is the absolute minimum since even if wealth drops to zero as a result of  interactions with other individuals within one fiscal cycle, Eq. (\ref{m2}), it is restored by the welfare system which in this version of the model increases wealth of everybody by the same amount equal to $\beta$ times the average wealth per capita, Eq. (\ref{m3}). The estimation of the minimum works well if the difference $1 - w_{min}$ is significantly larger than the standard deviation of the distribution of $\lambda$'s since then one can safely replace the random factors $\lambda_\tau$ by the mean, as we did in Eq. (\ref{fp}). The standard deviation is approximately equal to $\sigma$: $\sqrt{e^{\sigma^2}-1} \approx \sigma$ and $1-w_{min} \approx \frac{1-\beta}{\beta} g$ as follows from (\ref{abs_min}). Thus the estimation of the lowest level, Eq. (\ref{abs_min}), is valid if $\sigma \ll \frac{1-\beta}{\beta} g$. 
The system does not stay long at the absolute minimum, given by Eq. (\ref{abs_min}). While calculating the minimum we have assumed the maximal loss 
of wealth by the poorest individual, which in one fiscal period is $g w_{min}$. As follows from Eq. (\ref{tij_mf}) such a situation takes place only if there is a large wealth gap between wealth of the poorest individual and the next-to-poorest one, that is if $|J| (w_i - w_{min}) > g w_{min}$. Otherwise the wealth loss is smaller and the loss depends on the number of individuals belonging to the poorest class {\em i.e.} having a similar wealth as the poorest one. An individual $i$ belongs to this class if $|J| (w_i - w_{min}) \le g w_{min}$. The average gap for this group is estimated to be proportional to the standard deviation of $\lambda$'s that is $w_i-w_{min} \approx \sigma w_{min}$. Denoting the number of inviduals in the poorest class by $N_1$ one can estimate that the loss of wealth of the poorest individual is of order $(1-n_1) g w_{min} + n_1 \sigma w_{min}|J|$ where $n_1 = (N_1-1)/(N-1) \approx N_1/N$ is the fraction of all individuals in the poorest class. The first term estimates the outflow of wealth of the poorest individual to individuals from richer classes, and the second one to individuals from the poorest class. The fixed point equation takes the form 
\begin{equation}
w_{min} = (1-\beta) \left\{1 - (1-n_1) g - n_1 \sigma |J|  \right\} w_{min} + \beta  \ .
\end{equation}
It yields
\begin{equation}
w_{min}(n_1) = \frac{\beta}{\beta + g(1 -\beta)(1-n_1) + \sigma |J| n_1} 
\approx \frac{\beta}{\beta + g(1 -\beta)(1-n_1)} \ .
\label{min_n1}
\end{equation}
\begin{figure}
\includegraphics[width=0.5\textwidth]{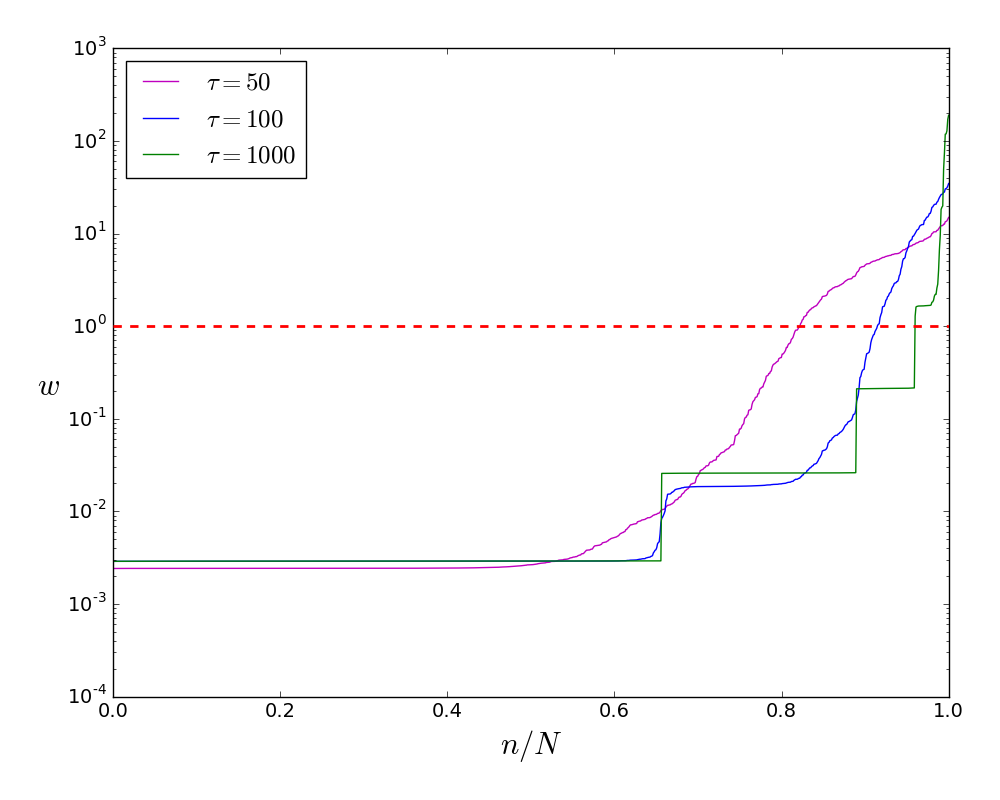}
\caption{Snapshots of wealth distribution at different times for the simulation shown in Fig. \ref{fig_evolution}. Each plot shows wealth versus normalised rank.
The normalised rank $n/N$ is on the horizontal axis. It is
ordered from the poorest individual, $n=1$, to the richest one, $n=N$.  Wealth is on the vertical axis, and it is in logarithmic scale. The dashed line in the plot is the equality line which corresponds to the uniform distribution {\em i.e.} with all agents having the same wealth. We used this configuration as an initial state in the simulations. The first snapshot (pink line) is taken after $\tau=50$ steps of evolution. One can see that the lowest level of the poorest class is below the asymptotic value (compare Fig \ref{fig_evolution}). The second snapshot (blue line) is taken after $\tau=100$ steps of evolution. One can see that the lowest level has reached the asymptotic value. The second poorest class has already formed which is seen as a step in the plot. The wealth level of this class is still below its asymptotic value. The third snapshot (green line) is taken after $\tau=1000$ steps of evolution. One can see a characteristic staircase structure which reflects the division of the population into wealth classes with almost flat distributions within classes and large wealth gaps between them. The length of each step is equal to the fraction of agents in the given class. \label{fig_rank}}
\end{figure}
The last approximation applies when $\sigma |J| \ll g$. Eq. (\ref{abs_min}) is a special case of Eq. (\ref{min_n1}) for $n_1=0$. The value of the lowest wealth level which is eventually reached during evolution depends on the fraction $n_1$ of individuals in the poorest class. For evolution shown in Fig. \ref{fig_evolution} there are $N_1=660$ individuals in the poorest class, so $n_1\approx 0.660$. The asymptotic wealth level estimated using Eq. (\ref{min_n1}) is shown in Fig. \ref{fig_evolution}. As one can see in the figure, the asymptotic level reached in the simulation conforms with the theoretical value. The size of the poorest class settles down when the poorest class gets separated from the remaining part of the population. Once the wealth gap between the poorest class and the next class is formed, it grows and becomes large as compared to the scale of statistical fluctuations. The class structure crystallises.
Actually, the system may split into many wealth classes. This effect is illustrated in Fig. \ref{fig_evolution}. Each group evolves towards its own final wealth level. The wealth levels of different classes are clearly separated from each other. The formation of classes can be seen also in Fig. \ref{fig_rank} which shows snapshots of sorted wealth data at different times.

The plot of sorted wealth has a characteristic step-like structure reflecting a division of the population into economic classes on different wealth levels $w_l$. The $y$-axis is in logarithmic scale. The separation between consecutive wealth levels $\log w_{l+1}-\log w_l$ is larger than the wealth dispersion within classes. The dispersion is of order $\sigma$. The height of steps between consecutive levels $\log w_{l+1}$ and $\log w_l$ can be estimated by the analysis of the flow term $T_{ij,\tau}$, Eq. (\ref{tij_mf}). The wealth flow between agents belonging to two neighbouring wealth classes $l$ and $l+1$ is $|J|(w_{l+1}-w_l)$ but not larger than $gw_l$. The class become separated when 
$|J|(w_{l+1}-w_l)=gw_l$. This gives an estimation of the step height: 
\begin{equation}
\log w_{l+1} - \log w_{l} = \log \frac{g+|J|}{|J|} .
\label{step_height}
\end{equation}
Again this estimation works as long as the scale of fluctuations is small. The last formula tells us that each step (in logarithmic scale) has roughly the same height. The consecutive wealth levels grow geometrically $w_{l+1} \sim a w_l$, where $a = (g+|J|)/|J|$. The wealth dispersion within each class is of order $\sigma$. One can use Eq. (\ref{step_height}) to estimate the number of different wealth classes $(\log w_{max} - \log w_{min}) /\log a$ where the minimal wealth is given by Eq. (\ref{min_n1}) and the maximal one by
\begin{equation}
w_{max}= \frac{\beta + g(1-\beta)}{\beta + \frac{1}{N-1} g(1-\beta)} \approx 
1 + g\frac{1-\beta}{\beta}
\label{wmax}
\end{equation} 
The last equation can be derived using a similar line of arguments as the one that led us to Eq. (\ref{abs_min}). The approximation in the last formula is valid when
the tax level is not too small $\beta \gg 1/N$. We see that in this case the wealth of the richest individual is a finite multiple of the average per capita. The situation changes when $\beta$ is close to zero in which case the upper limit $w_{max}$ grows with the system size $N$.
\begin{figure}
\includegraphics[width=0.7\textwidth]{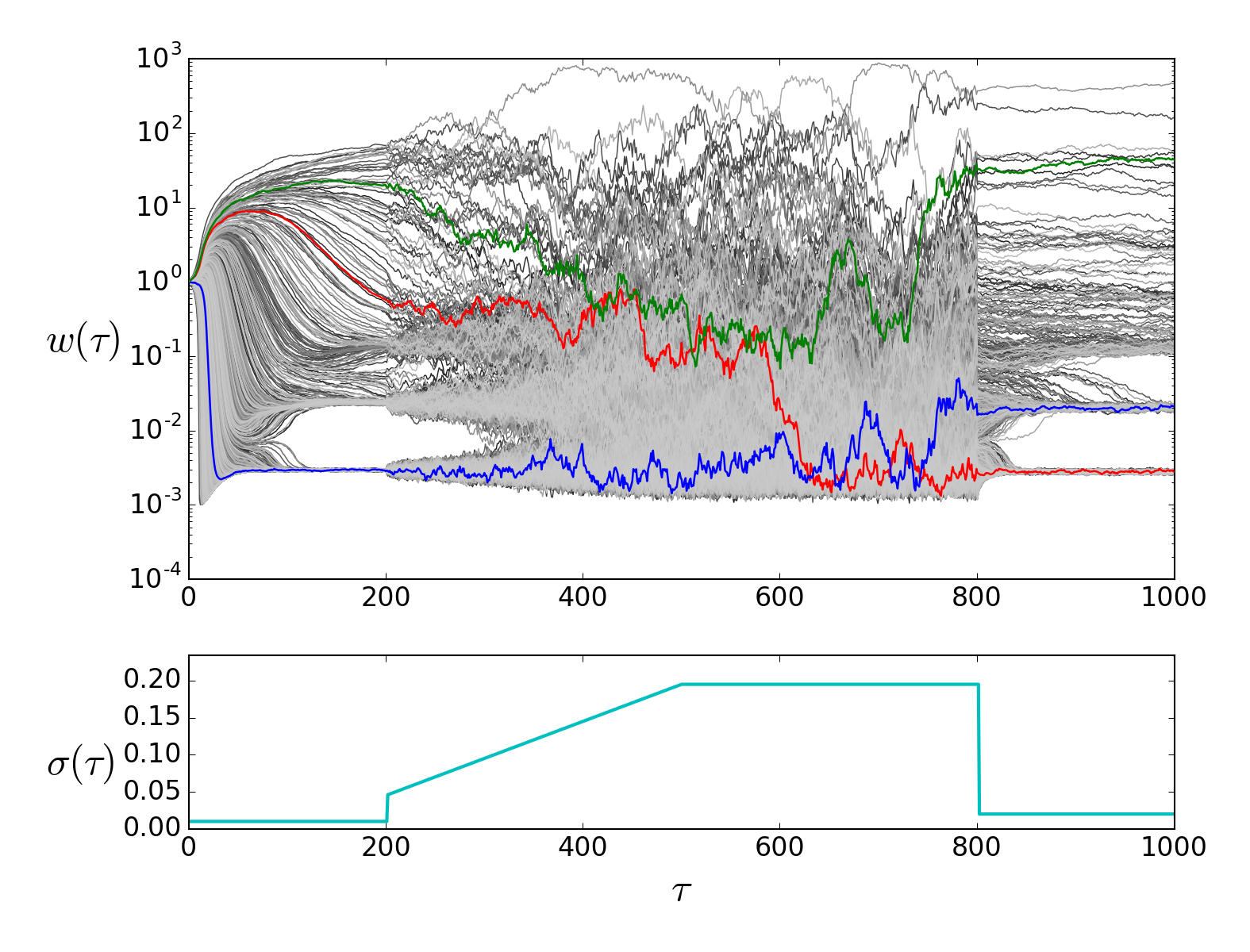}
\caption{Evolution of the system with varying $\sigma$. The $\sigma$-profile is shown at the bottom. For small $\sigma$ evolution is laminar. It becomes turbulent when $\sigma$ gets large. The red, green and blue lines in the figure represent evolution of wealth of three selected individuals: the blue curve - an individuals who was in the poorest class at the beginning and jumped to a middle class during the turbulent period (when $\sigma$ was large). The red line represents an individual who was rich at the beginning but he ended up in the lowest class. As long as $\sigma$ is small evolution is laminar and the individuals stay within their economic classes. The situation changes as soon as $\sigma$ gets larger. Wealth starts to fluctuate and the curves cross each other: poorer gets richer and vice versa. When $\sigma$ drops again the class structure crystallises. 
\label{fig_evol2}}
\end{figure}
As we mentioned for small $\sigma$ the evolution is laminar and the system is driven to a state consisting of separated wealth classes representing groups of individuals on a comparable wealth level, Fig. \ref{fig_evolution}). The histogram of the logarithm of wealth consists of separated peaks corresponding to different classes. The width of each peak is roughly proportional to $\sigma$. When the scale of fluctuations, $\sigma$, is increased the dispersion of each wealth class increases and the corresponding peaks get broader. When $\sigma$ gets even larger the peaks of neighbouring classes begin to overlap. Eventually they start to form a single broad distribution, see Fig. \ref{fig_sigma1}. 

Growing fluctuations smear the class structure. For large $\sigma$ individuals may move from one class to another. Moreover, the wealth classes are not sharply defined anymore. As the scale of fluctuations gets larger the evolution ceases to be laminar and becomes turbulent with many flips between classes. This is illustrated in Fig. \ref{fig_evol2}) where we show how the character of the evolution changes when $\sigma$ changes.


\begin{thebibliography}{99}
\bibitem{p} T. Piketty, {\em Capital in the XXI Century}, Harvard University Press, (2014);
\bibitem{sz} E. Saez and G. Zucman, {\em Wealth Inequality in the United States since 1913: Evidence from Capitalized Income Tax Data},
NBER Working Paper Series, Working Paper 20625, http://www.nber.org/papers/w20625 (2014);
\bibitem{p1} T. Piketty, {\em About Capital in the Twenty-First Century},
American Economic Review: Papers \& Proceedings  105(5): 48–53 (2015); http://dx.doi.org/10.1257/aer.p20151060
\bibitem{ox} Oxfam Briefing Paper, {\em An  economy for the 99\%}, January 2017, 
https://www.oxfam.org/
\bibitem{cs} Credit Suisse Research Institute, {\em Global Wealth Databook 2016}, November 2016; 
\bibitem{ds} J. B. Davis and A. E. Shorrocks, {\em The distribution of wealth}, 
Chapter 11 in {\em Handbook of Income Distribution}, ed.  Anthony B. Atkinson and F. Bourguignon, Elsevier, (2000);
\bibitem{bb} J. Benhabib and A. Bisin, {\em Skewed wealth distributions: theory and empirics}, NBER Working Paper Series, Working Paper 21924,
http://www.nber.org/papers/w21924 (2016);
\bibitem{obt} J. D. Ostry, A. Berg, C. G. Tsangarides, {\em Redistribution, Inequality and Growth}, IMF Discussion Note, SDN 14/02 (2014);
\bibitem{b} J.-P. Bouchaud,  J. Stat. Mech., (2015) P11011.
\bibitem{p2} V. Pareto, {\em Cours d'\'Economie Politique}, II, F. Rouge, Lausanne (1897);
\bibitem{s} P.A. Samuelson {\em A Fallacy in the Interpretation of the Pareto's Law of Alleged Constancy of Income Distribution}, Rivista Internazionale di Scienze Economiche
e Commerciali, 12, 246-250, (1965);
\bibitem{g} R. Gibrat, {\em Les In\'{e}galit\'{e}s \'{e}conomiques}, Paris, 1931;
\bibitem{ps} T. Piketty and E. Saez, {\em A theory of optimal capital taxation},
NBER Working Paper Series, Working Paper 17989, http://www.nber.org/papers/w17989 (2012);
\bibitem{cb} R. Cont, J.P.Bouchaud, {\em Herd behavior and aggregate fluctuation in financial market}, Macroeconomics Dynamics 4, (2000);
\bibitem{a} J. Angle, {\em The surplus theory of social stratification and the size 
distribution of personal wealth}, Social Forces 65, 293-326, (1986); 
\bibitem{szsl} E. Samanidou, E. Zschischang, D. Stauffer and T. Lux,
{\em Agent-based models of financial markets}, Reports on Progress in Physics 70, 409–450 (2007);
\bibitem{y} G.U. Yule, {\em A mathematical theory of evolution, based on the conclusions of Dr. J. C. Willis}, Philosophical Transactions of the Royal Society B 213, 21 (1925);
\bibitem{sim} H.A. Simon, {\em On a Class of Skew Distribution Functions}, Biometrika, 42, 425 (1955);
\bibitem{ba} A.L. Barabasi, R. Albert (1999), {\em Emergence of Scaling in Random Networks}, Science 286, 509 (1999);
\bibitem{bm} J.-P. Bouchaud and M. M\'{e}zard, {\em Wealth condensation in a
simple model of economy}, Physica A 282, 536 (2000);
\bibitem{d} F.J. Dyson, {\em A Brownian--Motion Model for the Eigenvalues of a Random Matrix}, Journal of Mathematical Physics 3, 1191 (1962);
\bibitem{f} P. J. Forrester, {\em Log-Gases and Random matrices}, Princeton University Press, 2010;
\bibitem{bbj} P. Bialas, Z. Burda, D. Johnston, {\em Condensation in the Backgammon model}, Nucl. Phys. B 493, 505 (1997);
\bibitem{bbbj} P. Bialas, L. Bogacz, Z. Burda, D. Johnston, {\em Finite size scaling of the balls in boxes model}, Nucl. Phys. B 575, 599 (2000);
\bibitem{xg} X. Gabaix, {\em Power Laws in Economics and Finance},
Annual Review of Economics, 1, 255 (2009);
\bibitem{dn} M. De Nardi, {\em Models of Wealth Inequality: A Survey} ,
Working Paper 21106, NBER (2015);
\bibitem{ap} A. Adamou, P. Ole, {\em Dynamics of Inequality}, Significance, 13, 32 
(2016);
\bibitem{bpa} Y. Berman, O. Peters, A. Adamou, {\em Far from equilibrium: Wealth reallocation in the United States}, arXiv:1605.05631, (2016);
\bibitem{bbs} Y. Berman, E. Ben-Jacob, Y. Shapira, {\em The Dynamics of Wealth Inequality and the Effect of Income Distribution}, PLOS ONE, 11, e0154196 (2016);
\bibitem{bfp} T. Blanchet, J. Fournier, T. Piketty, {\em Generalized Pareto Curves:
Theory and Applications}, WID.world working paper series 2017/3; 
\bibitem{ls} M. Levy, S. Solomon, {\em Power laws are logarithmic Boltzmann laws},
International Journal of Modern Physics C 7 , 595 (1996);
\bibitem{ikr} S. Ispolatov, P. L. Krapivsky, S. Redner, {\em Wealth Distributions in Models of Capital Exchange}, European Physical Journal B 2, 267, (1998); 
\bibitem{yr} V.M. Yakovenko, J.B. Rosser,  Jr. {\em Colloquium:
Statistical mechanics of money, wealth, and income}, Reviews of Modern Physics 81, 1703,  (2009);
\bibitem{bjn} Z. Burda, J. Jurkiewicz, M.A. Nowak, {\em Is Econophysics a Solid Science?},  Acta Physica Polonica B34, 87 (2003);
\bibitem{cccc} B. K. Chakrabarti, A. Chakraborti, S. R. Chakravarty, and A. Chatterjee, {\em Econophysics of Income and Wealth Distributions}, Cambridge University Press, 2013;
\bibitem{mhs} T. Ma, J.G. Holden, R.A. Serota, {\em Distribution of wealth in a network model of the economy}, Physica A 392, 2434 (2013);
\bibitem{ls1} Z. Liu, R.A. Serota, {\em Correlation and relaxation times for a stochastic process with a fat-tailed steady-state distribution}, Physica A 474, 301 (2017);
\bibitem{bss} Y. Berman, Y. Shapira, M. Schwartz, {\em A Fokker-Planck model for wealth inequality dynamics}, EPL 118, 3 (2017);
\bibitem{ls2} Z. Liu, R.A. Serota, {\em On absence of steady state in the Bouchaud–M\'{e}zard network model}, Physica A 491, 391 (2018);
\bibitem{gk} B.V. Gnedenko, A.N. Kolmogorov, {\em Limit distributions for sums of independent random variables}, Cambridge: Addison-Wesley (1954);
\bibitem{i} K. It\^{o}, {\em On stochastic differential equations}, Memoirs, American Mathematical Society 4, 1, (1944) ;
\bibitem{vk} N.G. Van Kampen, {\em Stochastic Processes in Physics and Chemistry (3rd Edition)}, Elsevier (2007) ;
\bibitem{k} H. Kesten, {\em Random Difference Equations and Renewal Theory for Products
of Random Matrices}, Acta Mathematica, 131, 207 (1973);
\bibitem{c} D.G. Champernowne, {\em A Model of Income Distribution,} Economic Journal,
63, 318, (1953);
\bibitem{ww} H.O.A. Wold, P. Whittle, {\em A Model Explaining the Pareto Distribution of
Wealth}, Econometrica 25, 591, (1957);
\bibitem{gr} I.S. Gradshteyn and I.M. Ryzhik, {\em Table of Integrals, Series, and Products}, Academic Press, 7th Ed., (2007); 
\bibitem{pyz} T. Piketty, L. Yang, G. Zucman, {\em
Capital Accumulation, Private Property and Rising Inequality in China, 1978-2015},
WID.world working paper series, No 2017/6, (2017);
\bibitem{bjjknpz} Z. Burda, et al, {\em Wealth condensation in Pareto macroeconomies},
Physical Review E 65, 026102 (2002);
\bibitem{b1} P. Ball, Nature, News020121-14, {\em Wealth spawns corruption,
Physicists are explaining how politics can create the super-rich}, doi:10.1038/news020121-14 (2002);

\end{thebibliography}
\end{document}